\documentclass[sigconf]{acmart}

\usepackage{longtable}
\usepackage{float}
\usepackage[most]{tcolorbox}

\usepackage{tikz}
\newcommand\mybox[2][]{\tikz[overlay]\node[fill=yellow!70,inner sep=2pt, anchor=text, rectangle, rounded corners=1mm,#1] {#2};\phantom{#2}}

\usepackage{csquotes}
\usepackage{todonotes}
\usepackage{array}
\usepackage{tabularray}
\usepackage{multirow}
\newcolumntype{C}[1]{>{\centering\let\newline\\\arraybackslash\hspace{0pt}}m{#1}}
\newcolumntype{P}[1]{>{\centering\arraybackslash}p{#1}}

\graphicspath{{figures/}}

\AtBeginDocument{%
  }

\copyrightyear{2025}
\acmYear{2025}
\setcopyright{cc}
\setcctype{by}
\acmConference[ASSETS '25]{The 27th International ACM SIGACCESS Conference on Computers and Accessibility}{October 26--29, 2025}{Denver, CO, USA}
\acmBooktitle{The 27th International ACM SIGACCESS Conference on Computers and Accessibility (ASSETS '25), October 26--29, 2025, Denver, CO, USA}
\acmDOI{10.1145/3663547.3746394}
\acmISBN{979-8-4007-0676-9/2025/10}

\begin{document}

\title[Design Opportunities for Just-in-Time OCD Intervention]{``It was Mentally Painful to Try and Stop'': Design Opportunities for Just-in-Time Interventions for People with Obsessive-Compulsive Disorder in the Real World}


\author{Ru Wang}
\email{ru.wang@wisc.edu}
\affiliation{%
  \institution{University of Wisconsin--Madison}
  \city{Madison}
  \state{WI}
  \country{USA}
}

\author{Kexin Zhang}
\email{kzhang284@wisc.edu}
\affiliation{%
  \institution{University of Wisconsin--Madison}
  \city{Madison}
  \state{WI}
  \country{USA}
}

\author{Yuqing Wang}
\email{ywang2776@wisc.edu}
\affiliation{%
  \institution{University of Wisconsin--Madison}
  \city{Madison}
  \state{WI}
  \country{USA}
}

\author{Keri Brown}
\email{keri@pureocdtherapy.com}
\affiliation{%
  \institution{Pure OCD Therapy}
  \city{Madison}
  \state{WI}
  \country{USA}
}

\author{Yuhang Zhao}
\email{yuhang.zhao@cs.wisc.edu}
\affiliation{%
  \institution{University of Wisconsin--Madison}
  \city{Madison}
  \state{WI}
  \country{USA}
}

\renewcommand{\shortauthors}{Wang et al.}

\begin{abstract}

Obsessive-compulsive disorder (OCD) is a mental health condition that significantly impacts people's quality of life. While evidence-based therapies such as exposure and response prevention (ERP) can be effective, managing OCD symptoms in everyday life---an essential part of treatment and independent living---remains challenging due to fear confrontation and lack of appropriate support. To better understand the challenges and needs in OCD self-management, we conducted interviews with 10 participants with diverse OCD conditions and seven therapists specializing in OCD treatment. Through these interviews, we explored the characteristics of participants' triggers and how they shaped their compulsions, and uncovered key coping strategies across different stages of OCD episodes. Our findings highlight critical gaps between OCD self-management needs and currently available support. Building on these insights, we propose design opportunities for just-in-time self-management technologies for OCD, including personalized symptom tracking, just-in-time interventions, and support for OCD-specific privacy and social needs---through technology and beyond.

\end{abstract}





\maketitle

\section{Introduction}
Obsessive-Compulsive Disorder (OCD) is a mental health disorder affecting approximately 2\% of world population \cite{sasson1997epidemiology}. People with OCD experience uncontrollable, recurring thoughts (\textit{obsessions}) usually triggered by specific events (\textit{triggers}), and engage in repetitive behaviors (\textit{compulsions}) to relieve the anxiety caused by obsessions \cite{nimh}. 
Contrary to the common misconception that OCD is simply `being clean' or `focused on order' \cite{spencer2021isn}, 
OCD symptoms are highly diverse \cite{hirschtritt2017obsessive} and can significantly affect quality of life especially when compulsions become severe enough to disrupt daily activities \cite{macy2013quality, eisen2006impact, subramaniam2013quality}.

Coping with OCD symptoms independently in daily life is crucial but challenging as real-life triggers can be too overwhelming to confront without suitable guidance \cite{lind2013technological, hezel2019exposure, wheaton2021homework, ojalehto2020adherence}. Many people with OCD seek professional support through cognitive behavioral therapy (CBT) to receive help in identifying and changing obsessive thoughts and behavior patterns \cite{beck2011cognitive}. 
Despite its effectiveness \cite{yule2017cognitive, blakey2017exposure, hirschtritt2017obsessive}, over 50\% of people with OCD lack access to the therapy treatment \cite{garcia2014factors, ruscio2010epidemiology, chong2012treatment, mayerovitch2003treatment}. Moreover, many who complete the treatment continue to experience residual symptoms and struggle to apply the skills learned in therapy to their daily lives \cite{eddy2004multidimensional}. These challenges highlight the needs for effective just-in-time self-management technology that supports people with OCD in real life beyond traditional therapy.

Recent advances in AI and sensing techniques have shown great potential for just-in-time OCD interventions. Through intelligent scene understanding and behavior recognition \cite{xu2024survey, xie2024large, bordes2024introduction, liu2024human}, such technology can potentially detect triggering events and compulsive behaviors and generate timely, in-situ interventions to help people with OCD address obsessions and resist compulsions during OCD flare-ups \cite{burchard2022washspot,olbrich2016smartphone}. 
Despite the potential, HCI research on just-in-time intervention for OCD remains limited and primarily focuses on specific symptoms, for example, 
recognizing excessive checking behaviors when a short moving distance over an extended period of time is detected based on GPS location and providing audio alarm to remind the users \cite{olbrich2016smartphone}. However, OCD is way more complex and heterogeneous with diverse triggers and symptoms \cite{hirschtritt2017obsessive}. Timely research is needed to deeply understand the diverse triggering experiences of people with OCD, uncover their challenges and needs, and identify suitable intervention feedback, thus inspiring effective, generalizable technology design.

To fill this gap, we contribute the first interview study with ten people with OCD and seven OCD professionals to understand the diverse triggers and experiences of OCD, the persisting challenges faced by OCD individuals, real-life coping strategies, and desired (or recommended) technology for just-in-time interventions. Our study cover the perspectives of both stakeholders to identify the agreement and tension between people with OCD and the professionals, thus deriving most suitable intervention design guidelines that not only fulfill user needs but also align with professional practices.

Our study revealed unique findings towards OCD users' experiences and preferences, as well as their tension with the professionals. Specifically, we found that instead of the trigger type in itself (e.g., a trash can), it was the different properties of triggers (e.g., fullness of the trashcan, whether there is a lid) that predominantly affect user experience. Moreover, besides certain consistency, we identified conflicts between the coping strategies adopted by people with OCD and those taught by OCD therapists. For example, the relaxing techniques (e.g., breathing practice) commonly used by OCD users in an OCD episode were not always recommended by the professionals. Our research highlights the gaps in OCD self-management despite currently available resources and derives technology design implications from trigger and compulsion detection, intervention feedback design, and ethics and privacy perspectives. Beyond just-in-time interventions in real life, some of our implications can also benefit OCD therapy, such as supporting out-of-session practice.

\section{Related Work} 

\label{related work}
\subsection{Background of OCD}
\label{def ocd}
Obsessive-Compulsive Disorder
occurs when a person is caught in a loop of obsessions and compulsions \cite{american2013diagnostic}. \textit{Obsessions} are repetitive, intrusive, and unwanted thoughts, urges, or mental images caused by certain triggers and induce distress feelings such as fear, anxiety, and disgust \cite{apa}. 
\textit{Compulsions} are repetitive overt behaviors or covert mental rituals (e.g., repetitively reviewing events mentally to prevent harm) performed to relieve distress caused by obsessions \cite{olatunji2013cognitive}; avoidance of people, places, or situations that trigger obsessions can also be a manifestation of compulsion \cite{apa}. 
People with OCD might feel temporary relief by conducting compulsive behaviors (e.g., excessive hand washing), but in no way doing these compulsions can bring them pleasure \cite{nimh}.
In severe situations, obsessions and compulsions can take up hours of a person’s day and can affect family and social relationships \cite{koran2000quality}. As a result, OCD poses barriers to both professional and personal life, leading to diminished quality of life \cite{bobes2001quality, subramaniam2013quality, macy2013quality}. 

OCD is highly heterogeneous. 
Yale-Brown Obsessive Compulsive Scale (Y-BOCS) Symptom Checklist \cite{goodman1989yale} organizes different OCD symptoms into eight obsession categories and seven compulsion categories \cite{goodman1989yale, mckay2004critical}. Common obsession themes include fear of contamination, persistent doubting, violent or sexual intrusive thoughts. 
In response to these obsessions, the compulsive behaviors can be excessive hand washing, repeated checking (e.g., whether the oven is off), and repeated neutralizing thoughts (e.g., ``I'm not a bad person''), respectively \cite{hirschtritt2017obsessive}. However, \textbf{unlike obsessions and compulsions, the diversity of triggers that cause obsessions and compulsions has been overlooked in OCD research}. Existing studies focus on generic triggers for specific symptoms \cite{mataix2009maudsley} or describe OCD episodes without clearly distinguishing triggers \cite{knapton2016experiences}. Since triggers are highly personalized and important in characterizing OCD symptoms \cite{baioui2013neural}, it is crucial to understand and categorize triggers by exploring OCD people's personal experiences.

To combat OCD, cognitive behavioral therapy (CBT) is the most effective method \cite{hirschtritt2017obsessive}, focusing on identifying and changing people's unwanted thinking and behavioral patterns \cite{beck2011cognitive}. 
CBT for OCD includes different forms, such as exposure and response prevention (ERP) therapy, cognitive therapy (CT), and acceptance and commitment therapy (ACT) \cite{bjorgvinsson2007obsessive}.
ERP helps people with OCD gradually face their OCD triggers while refraining from compulsions, aiming for behavioral change through repeated exposure to those triggers \cite{blakey2017exposure, bjorgvinsson2007obsessive, hirschtritt2017obsessive}. Cognitive Therapy (CT) focuses on identifying and restructuring dysfunctional beliefs that lead to obsessions, helping individuals reevaluate perceived catastrophic outcome \cite{yule2017cognitive}. Acceptance and Commitment Therapy (ACT), in contrast, encourages acceptance of obsessive thoughts without engaging in compulsions and promotes value-driven behavior to reduce the impact of OCD symptoms \cite{smith2017acceptance, smith2017acceptance, philip2021acceptance, evey2023systematic}. CBT has proven to be effective in addressing OCD symptoms from both obsession and compulsion aspects. However, it is not available 24/7 whenever people need support to manage their issues \cite{blakey2017exposure}. The homework between sessions (e.g., symptom self-monitoring, self-guided exposure practices)---as an important part of CBT---could be difficult for clients to comply, affecting the outcome of therapy \cite{wheaton2021homework}. Despite the drawbacks, the strategies used in these therapies can potentially bring insights for just-in-time interventions. Our research thus seize this opportunity to conduct studies with people with OCD to understand their experiences and coping strategies, triangulated with professionals' perspectives, to inspire real-world intervention technologies beyond conventional therapy.

\subsection{Technology for OCD Symptom Self-Management} 
To overcome the challenges caused by OCD in real life, different forms of technologies have been developed and evaluated. Existing efforts have mainly focused on symptom and behavior monitoring \cite{khan2011use, frank2023wearable} and proactive skill building \cite{khan2011use, ferreri2019new, hong2020multimedia, boisseau2017app, abramowitz2010assessment}, but limited work has focused on just-in-time interventions for specific OCD symptoms \cite{burchard2022washspot, olbrich2016smartphone}. We discuss prior work on OCD self-management technology below.

\subsubsection{Symptom monitoring}
Symptom monitoring is an effective approach to supporting mental health self-management, such as anxiety disorder and depression \cite{kruzan2023perceived}, due to its potential improvement in symptom (e.g., thoughts, behavior) awareness \cite{kruzan2023perceived}. Specifically in the context of OCD, prior work has explored two ways of self-monitoring, (1) self-report, and (2) passive sensing.

The most common self-report approach is via Ecological Momentary Assessment (EMA), which periodically prompts the users to track and assess their own symptoms in their natural environments \cite{frank2023wearable, brown2020implementation, tilley2014clinical, rupp2020comparing}. While this method has demonstrated high acceptability, practicability, and representativeness \cite{frank2023wearable}, research has shown that overly relying on self-report approach can result in noisy and missing data \cite{frank2023wearable}.

In contrast, passive sensing enables rich data collection without burdening users. Researchers have investigated passive symptom monitoring using various sensors \cite{thierfelder2022multimodal, alfano2011objective, drummond2012should, coles2020sleep, cox2022delayed, lonfeldt2023predicting}. For example, Thierfelder et al. \cite{thierfelder2022multimodal} monitored OCD-induced stress by measuring heart rate, hand activity and gaze fixation, and demonstrated the possibility of using multimodal sensory data to detect stress and anxiety. 
Other studies have focused on monitoring different aspects of OCD symptoms, including compulsion (e.g., excessive hand-washing) detection using inertial motion sensor data on a smartwatch \cite{wahl2022automatic, burchard2022washspot}, OCD episode tracking using blood volume pulse and skin temperature \cite{lonfeldt2023predicting, olesen2023predicting}, and sleep quality measurement using actigraphy \cite{alfano2011objective, drummond2012should, coles2020sleep, cox2022delayed}. 
While symptom monitoring can improve symptom awareness, prior work has primarily focused on monitoring users' internal states instead of contextual information (e.g., potential triggers in the environment), which is important in characterizing OCD symptoms \cite{llorens2021context}. In addition, without proper knowledge and coping skills for symptom management, the effectiveness of symptom monitoring can be further limited. 


\subsubsection{Proactive Skill Building}

To support skill building outside therapy sessions, virtual therapy applications based on desktop or smartphone have been developed \cite{ferreri2019new, hong2020multimedia, boisseau2017app, abramowitz2010assessment}. For example, \textit{LiveOCDFree} is an app that guides the user to set up and conduct self-administered ERP therapy by creating an exposure hierarchy and identifying specific triggers to practice (e.g., touching doorknobs). The user is able to set reminders for ERP practice and monitor their progress. An evaluation with 21 participants demonstrated significant improvement in OCD and anxiety symptoms after using the app \cite{boisseau2017app}. Similarly, \textit{NOCD} supported
self-administered ERP between therapy sessions by incorporating ACT strategies. A study with 2,069 participants showed that it effectively reduced OCD symptoms \cite{feusner2021initial}. 
Beyond conventional applications, recent advances in large language models (LLMs) further presents new opportunities in delivering personalized mental well-being support
\cite{guo2024soullmate, kian2024can, nie2024llm, iftikhar2024therapy}. For example, Kian et al. \cite{kian2024can} developed an LLM-powered assistive robot to deliver out-of-session CBT exercises for users with depression and achieved comparable effectiveness to traditional methods. 

All above therapy simulation technologies require users to proactively practice to build up coping skills. However, triggers in real world could be much more unpredictable and intense, which can still be hard to manage even with certain skills established \cite{eddy2004multidimensional}. 
In contrast, targeted, just-in-time interventions that are adaptive to real world events \cite{kim2024opportunities} can potentially be more effective in real life. 

\subsubsection{Just-in-time Interventions}
As opposed to pre-scheduled therapy sessions, the increasingly powerful mobile and sensing technologies allow for more flexible and intelligent mental well-being interventions. Such systems can recognize and adapt to users' internal mental states and physical context in daily life and provide just-in-time support---an intervention design that delivers the right type and amount of support, adaptive to the timing and context when the user needs it. \cite{nahum2018just}.
Just-in-time interventions have been applied to various mental health conditions, such as substance use disorder \cite{gustafson2014smartphone}, depression \cite{burns2011harnessing, he2024affective, zhao2023affective, costa2016emotioncheck}, schizophrenia \cite{ben2014feasibility, ben2013development}, and excessive phone use \cite{xu2022typeout}
For example, \textit{ACHESS} \cite{gustafson2014smartphone} is a mobile app that tracks the GPS location of a user with alcohol use disorders and provides text-based reminder when they go to high-risk locations (e.g., a bar). An evaluation with 349 participants shows that ACHESS significantly reduced risky drinking days.
Zhao et al. \cite{zhao2023affective} developed a wearable device that effectively reduced anxiety during high-stress situations through affective touch---a tactile intervention that renders a soft stroking sensation on the user’s forearm. 
Moreover, Xu et al. \cite{xu2022typeout} designed \textit{TypeOut}, which integrates value-based self-affirmation messages with a typing-based app unlock process, significantly reducing app usage and screen time. 

Despite the exploration of just-in-time interventions for other mental health conditions, the work for OCD remains nascent. 
The only work we know is from Olbrich et al. \cite{olbrich2016smartphone}, who develops a smartphone app \textit{Geo-Feedback} that actively detects users' excessive checking behaviors indicated by short moving distance over an extended period of time via GPS location and evokes an alarm when such excessive behaviors are detected. A case study with one user with severe OCD indicates the effectiveness of the app. 
While demonstrating the initial success of just-in-time interventions for people with OCD, this work focuses only on one specific OCD condition.
More research is needed to understand the diverse, complex nature of OCD as well as user needs, thus deriving more generalizable guidelines for just-in-time intervention technology (including both trigger detection and intervention design) to support various OCD types in real life. 
\section{Method}
To better understand how OCD (e.g., obsessions, compulsions, triggers) impacts people's daily life and to identify effective just-in-time intervention strategies, we conducted in-depth interviews with 10 people with OCD (P1-P10) and seven OCD therapists (T1-T7). Our goal was to distill the triggering events, strategies, and challenges of OCD self-management in real life, ultimately informing the design of just-in-time intervention technologies.

\begin{table*}[h!]
\footnotesize 
\centering
\begin{tabular}{C{0.4cm}C{0.9cm}C{2cm}C{2cm}C{5cm}}
\toprule
\textbf{ID} & \textbf{Age/ Gender} 
& \textbf{Severity} & \textbf{Diagnosed Time} 
&  \textbf{OCD Types} 
\\ \hline
P1 & 19/F
& Mild-to-moderate  & 1.5 years ago   
& Contamination OCD, responsibility OCD   \\ \midrule 
P2 & 21/F
& Mild  & 1.5 years ago   
& Harm OCD, sexual OCD \\ \midrule
P3 & 25/F 
& Moderate  & 16 years ago 
& Contamination OCD  \\ \midrule
P4 & 22/F 
& Moderate  & 3 years ago 
& Just-right OCD
\\ \midrule
P5 & 47/F 
& Mild  & 25 years ago 
& Responsibility OCD \\ \midrule
P6 & 25/F 
& Mild  & 6 years ago 
& Contamination OCD, responsibility OCD\\ \midrule
P7 & 25/F 
& Moderate  & 1 year ago  
& Contamination OCD, responsibility OCD\\ \midrule
P8 & 33/M 
& Moderate  & 6 years ago 
& Contamination OCD, responsibility OCD\\ \midrule
P9 & 33/M 
& Moderate  & 11 years ago 
& Responsibility OCD   \\ \midrule
P10 & 35/M 
& Moderate  & 6 months ago 
& Just-right OCD \\

 \bottomrule
\end{tabular}
\caption{Demographic information and basic obsession types of the 10 participants living with OCD. }
\label{tab:demo_OCD}
\end{table*}
\begin{table*}[h!]
    \scriptsize 
    \centering
    
    \begin{tabular}{C{0.5cm}C{0.9cm}C{3cm}C{1.5cm}C{3cm}C{2cm}C{1.5cm}}
    \toprule
      \textbf{ID}&  \textbf{Age/ Gender}&  \textbf{Certificate}&  \textbf{Years of {\newline} Experience}&  \textbf{Specialization}&   \textbf{OCD Clients {\newline} Severity Level}&   \textbf{Practice Modality} \\
     \midrule 
   
T1 & 46/F & Licensed Clinical Social Worker & 25 years & OCD, anxiety disorders, eating disorders, PTSD, depression, substance use disorder & Mild to severe & Hybrid \\
     \midrule 
T2 & 47/F & Clinical Psychologist & 14 years & OCD, axiety disorder & Mostly moderate to severe       & Virtual \\
     \midrule 
T3 & 36/M & License Professional Counselor & 9 years & OCD, anxiety disorder, phobias, health anxiety & Mild to severe, mostly moderate & Virtual \\
     \midrule 
T4 & 60/F & Licensed Clinical Social Worker & 13 years & OCD, anxiety, depression & Mostly moderate & Hybrid \\
     \midrule 
T5 & 51/F & Professional Counselor Training License (LPC-IT) & 2 years and 4 months & OCD, anxiety disorder, depression & Mild to severe, mostly moderate & Hybrid \\
     \midrule 
T6 & 38/M & License Professional Counselor & 17 years & OCD, anxiety disorders, eating disorders, PTSD & Mild to severe & Hybrid \\
     \midrule 
T7 & 29/F & License Professional Counselor & 2 years & OCD, anxiety disorder & Moderate to severe& Virtual\\

     \bottomrule
    \end{tabular}
    \caption{Demographic information and professional background of the seven OCD professionals.}
    \label{tab:demo_therapists}
\end{table*}

\subsection{Participants}
Our study involved two key stakeholders in the OCD domain---individuals who experience OCD and professionals who specialize in OCD treatment---to develop a comprehensive understanding of the current practices and challenges, ensuring that our design insights would integrate both lived experience of people with OCD and professional recommendations. 

\subsubsection{People with OCD}
We recruited 10 participants with OCD (7 female and 3 male) whose ages ranged from 19 to 47 ($Mean$ = 28.5, $SD$ = 8.5). Participants' demographic information and OCD types are detailed in Table \ref{tab:demo_OCD}. While our participant gender coverage skewed towards women, it was consistent with the higher prevalence of OCD in female adults than male adults \cite{mathes2019epidemiological}. 
We recruited participants from online OCD support groups and the research email service at our university. 
A participant was eligible if they were (1) at least 18 years old, (2) diagnosed with mild to moderate OCD, and (3) were willing to disclose their OCD symptoms with us. We did not recruit people experiencing severe OCD at the time of the study because our interview questions can be triggering without adequate ability to manage symptoms independently. Participants were compensated for \$20 per hour. 

Our participants demonstrated a wide range of OCD types (Table \ref{tab:demo_OCD}), covering most common obsessions and compulsions in the Y-BOCS Symptom Checklist \cite{goodman1989yale}. Six participants (P1, P5, P6-9) had responsibility OCD, whose obsessions were around not being responsible enough or fear of making mistakes; their common compulsions included checking if the door is properly locked. Five participants (P1, P3, P6-P8) had contamination OCD (e.g., dirt, germ) with common compulsions being excessive hand washing and cleaning. Two (P4, P10) had just-right OCD who were obsessed with symmetry or specific position of objects with corresponding compulsions of fixing asymmetry (P4) and objects that are not where they should be (P10). P2 experienced harm OCD, having intrusive thoughts about harm occurring to herself and people around her with compulsion of ruminating about the negative thoughts and avoiding the triggers. She also experienced sex-related obsessive thoughts (sexual OCD) when interacting with men and avoided looking at them to ease her distress. Consistent with the distribution of different OCD subtypes \cite{ball1996symptom}, more than half of our participants had contamination OCD and responsibility OCD. Table \ref{tab:ocd_symptoms} detailed participants' OCD subtypes.

\subsubsection{OCD Professionals}
We recruited seven OCD professionals (5 female, 2 male). Their ages ranged from 29 to 60 ($Mean$ = 43.9, $SD$ = 10.4). All participants had extensive experience working as mental health providers (at least two years), with four (T1, T2, T4, T6) having more than 10 years of experiences. 
Participants had experiences in treating OCD across various severity levels (from mild to severe) and in different practice modalities (both hybrid and virtual). 
Table \ref{tab:demo_therapists} shows participants' detailed information. 

We recruit OCD professionals by sending emails to local OCD therapists listed on the International OCD Foundation, a nonprofit organization providing resources and education for people with OCD \cite{iocdf}. 
Participants were eligible if they were over 18 and had experience conducting OCD related therapy. We limited the recruitment to individuals who spoke English and located in the United States. Each participant received \$40/hour as compensation.

\subsection{Procedure}
We conducted single-session semi-structured interviews with both participants with OCD and OCD professionals via Zoom. Each session lasted for about 90 minutes. Our interview focused on different perspectives between the OCD individuals and the professionals. 

For individuals with OCD, we focused on their unique triggers, experiences, challenges, and coping strategies in daily life. Specifically, we first asked about their demographic information (e.g., age, gender identity) and diagnosis of OCD. Then we asked participants about their detailed personal experience with OCD---we prompted them about their obsessions, triggers, and compulsions in different living environments (e.g., home, workplace, public spaces). For triggers, we specifically asked for fine-grained details with a goal of identifying the characteristics and rationales of the triggers. 
We then asked participants about their strategies and technology use to manage their OCD symptoms in daily life, either recommended by therapists or developed by themselves. We also asked about their therapy experiences and use of other resources to assist with symptom management. 
Finally, participants discussed ideas about desired real-life interventions for symptom self-management. 

For OCD professionals, we focused on their strategies and practices in OCD therapy sessions to potentially inspire intervention design. We first asked about professionals' demographic information, certifications, and experience as a mental health therapy provider. We then asked about the types of therapy they conducted, the overall therapy process (e.g., treatment stages), the strategies and technologies used at different treatment stages, and how they evaluated client progress.
In addition to in-session practices, we asked therapists about out-of-session strategies they recommend, such as assigned homework and external support. We then had them discuss their current challenges in therapy practices. Lastly, we asked therapists about their familiarity with and opinions on emerging technologies, such as phone-based self-administered therapy. When therapists were unfamiliar with certain technologies (e.g., virtual reality), we provided a brief, neutral description to ensure informed responses and minimize misconceptions.

\subsection{Analysis}
We audio recorded and transcribed all interviews via an online transcription service (otter.ai\footnote{https://otter.ai}). One researcher manually cleaned all transcripts and corrected transcribing errors by checking the recorded videos. We analyzed the data using thematic analysis \cite{clarke2015thematic, braun2006using}. We selected two representative transcripts from OCD participants and two from OCD professionals as samples to develop initial codebooks. Two researchers coded the samples independently with open coding. Then, they discussed and reconciled their codes to resolve any differences and developed two initial codebooks upon agreement, one for participants with OCD and one for OCD professionals. One researcher then coded the remaining transcripts from the two groups of participants based on the codebooks, respectively.
If a new code was constructed, the two researchers discussed on the code and updated the codebooks upon agreement. 

We first derived themes for OCD participants and OCD professionals separately, where we used a hybrid of inductive and deductive approaches, a method commonly adopted in qualitative research \cite{oewel2024approaches, swain2018hybrid, braun2006using}. 
We started with high-level themes based on fundamental concepts of OCD triggers, compulsions, therapy treatment, and self-management identified in prior work \cite{wu2017insight, knapton2016experiences, blakey2017exposure, wu2017insight, yule2017cognitive} using a deductive approach \cite{crabtree1992template}. We then used an inductive approach \cite{boyatzis1998transforming} to generate sub-themes and identify new themes by grouping related codes via affinity diagramming \cite{clarke2015thematic, braun2006using} for each participant group.
Finally, we generated final themes and sub-themes across participants by comparing and combining themes from the two groups. Two researchers cross-referenced the original data, the two codebooks, and themes to make final adjustment, ensuring all codes were grouped in the correct themes.
Our analysis resulted in over 700 codes and 20 themes and sub-themes.

\subsection{Ethical Considerations \& Reflexivity Statement}
Our interviews involved discussing participants' OCD symptoms, which could cause mental distress. To provide a safe and comfortable environment for OCD participants, we consulted a clinical psychologist (a co-author) with experience treating OCD to ensure appropriate language was used in the interview.
OCD participants were informed that they could skip any uncomfortable questions and turn off their camera if preferred, and they could exit the study anytime when needed. 

We recognize that researchers' identities and background may influence our data interpretation. The primary researcher identifies as a person with OCD. His background and lived experience with OCD enabled us to develop appropriate questioning routes, better empathize with participants, and interpret their responses via a unique lens. Moreover, the research team comprises diverse expertise, including HCI, accessibility, intelligent system design, and OCD therapy, enabling us to better contextualize the study data in existing technology and mental health literature, thus deriving suitable technology implications to support people with OCD.

We interviewed both people with OCD and therapists who treat OCD to gather diverse perspectives on symptom self-management strategies and needs. While therapists offer insights grounded in treatment principles and their experience with clients, we acknowledge their positionality as mental health professionals rather than individuals with lived experience. Their interpretations may not fully capture, and may at times diverge from, how people with OCD understand and relate to their experiences. As such, our analysis prioritized the narratives of people with OCD and used therapist perspectives as a complementary lens. When tensions between the two groups emerged, we present them transparently without privileging therapists’ views.
\section{Findings}
Our study findings go beyond the conventional clinical view of OCD and takes a unique sociotechnial lens to understand the OCD individuals' unique triggers, compulsions, and coping strategies as well as highlighting the overlaps and contrast between the perspectives of people with OCD and the professionals. We detail our findings below.
\subsection{Redefining OCD Triggers}
\label{props of triggers}
We first investigated the OCD triggers to determine \textit{\textbf{what triggers to recognize}} and \textit{\textbf{how to recognize triggers}} for just-in-time interventions for OCD. 
We highlight that triggers go beyond ``a common list'' followed by prior work \cite{mataix2009maudsley} but include various types in different forms. Moreover, it is not the trigger type (e.g., a trashcan) in itself but rather its properties (e.g., fullness of the trashcan) that can significantly affect an OCD individual's experiences.


\subsubsection{Triggers can be anything} 
Unlike prior work that limits OCD triggers to a set of "common triggers" based on basic OCD themes (e.g., contamination, harm) \cite{mataix2009maudsley}, our findings show that OCD triggers are highly individualized.
While most triggers reported by OCD participants are specific \textit{physical objects} (i.e., triggering objects), such as trash cans for contamination OCD (P1) or knives for harm OCD (P2), participants also revealed less recognized trigger forms, including \textit{contexts}, \textit{sounds}, and even \textit{thoughts}.

We found that OCD can be triggered by participants' \textit{physical context}, such as specific locations or times (P1, P3, P7-9). For instance, participants with contamination OCD (P1, P3, P8) were triggered every time they returned home, since germs brought from the outside can contaminate their home environment. 
Interestingly, weather can also be a trigger. For example, P7 indicated her fear of hitting a pedestrian emerged with low visibility on the road during rain or snow. 
OCD triggers can also be \textit{audible} (P3, P6-8, P10). For example, coughing sound can immediately trigger P3's contamination OCD, and accident-related keywords in a conversation would trigger P8 to check his own apartment for safety.
In addition, some participants reported \textit{mental} triggers (P5), such as unspecified worries, without clear links to physical objects. Moreover, OCD episodes could sometimes occur even without identifiable triggers (P1, P4, P6, P10).

\mybox{\textit{Inspiration}:}
The diverse forms of triggers reveal the difficulty of real-time trigger recognition and highlight the needs for the use of multi-modal sensory data in trigger identification.

\subsubsection{Trigger Intensity: Properties over Objects} Beyond the trigger type, the intensity of triggers (i.e., the mental distress induced by a certain trigger) plays an important role in determining whether an individual would be triggered. 
Our OCD professionals emphasized the importance of asking clients not just about the triggers themselves, but also about the factors that influence their intensity (T2, T3, T5). Specifically, the OCD professionals often work with the clients to create a \textit{trigger hierarchy}---a ranked list of OCD triggers ordered based on their intensity. However, this process usually follows a specific template by asking about clients' experience with a checklist of ``common triggers'' (e.g., Y-BOCS \cite{goodman1989yale}) and then asking them to rate their distress caused by a specific trigger (T1-T2, T4-T7) or their urge to performance compulsion (T3). While some professionals allowed their clients to add more diverse triggers by logging their encounters outside the therapy session (T1–T5, T7), we found that there was a lack of comprehensive method in clinic to identify all possible triggers. Moreover, the method of evaluating trigger intensity was completely based on client's subject ratings, without systematic characterization of the factors that affect trigger intensity.  


Despite the lack of clinical standard, our study data from OCD individuals revealed that, for physical triggers, instead of the object type in itself, it was often the specific properties of the object that determined the trigger intensity, 
leading to varying levels of anxiety and compulsions. 
Below, we summarize key properties that affect trigger intensity from the experiences of different participants with OCD. 

\textbf{\textit{Sensory Attributes.}}
Participants reported multiple sensory attributes that can alter the intensity of a trigger. 
Visual attributes such as \textit{color} can affect people's perception of a trigger (P3, P7). 
For example, P7's contamination OCD were triggered when the dirt presents high contrast against its background, such as coffee grind on a white countertop. While the same amount of white salt (also considered as dirt) with low contrast to the countertop would not be as triggering. 
In addition to color, some specific \textit{shapes} can also be more triggering than others, such as sharp edges. P2 felt more uncomfortable when interacting with objects that have sharp tips in her home environment, including knives with sharp blade, pencils with sharp tips, and the newel posts with sharp corners on a staircase. 
Moreover, \textit{tactile attributes} of an object could be another factor that affected trigger intensity (P1, P7). For instance, the sticky texture of sauce (e.g., ketchup) caused more contamination concerns to P1 and P7 than solid food.

\textbf{\textit{Affordances.}}
We found triggering objects that afforded direct contact with the user can cause higher level of distress to participants with contamination OCD (P1, P3, P6-8). For example, door handles in public spaces and handrails in transportation are common triggers because it is meant to be touched using hands by many people, thus considered dirtier than surfaces not requiring direct contact (e.g., automatic doors). P1 mentioned trash can as one of her OCD triggers, but the ones with a lid and handle were more triggering for the same aforementioned reason. 

\textbf{\textit{Spatial Attributes.}}
Spatial attributes of a triggering object is another property that determines trigger intensity. First, the \textit{quantity} of triggering objects can affect the intensity of contamination OCD. 
P3 tended to avoid occasions with big crowds of people more than others due to concerns about contagious diseases. Second, the \textit{proximity} of a trigger to a person with OCD is another factor that alters trigger intensity. According to P2, a knife in hand caused more harm-related intrusive thoughts to her than a knife on the counter top. 
Interestingly, trigger intensity can depend on the \textit{pose} and \textit{orientation} of a physical trigger. P2's sexual intrusive thoughts were more likely triggered when a man was in a position that ``resembled those in her intrusive thoughts.'' 

\textbf{\textit{Familiarity.}}
Trigger intensity varied depending on participants' familiarity with their triggers. For participants with harm OCD (P2), familiar triggers can be more intense. For example, P2 experienced more intrusive thoughts about accidentally hurting people when around family or friends rather than strangers.
Conversely, participants with contamination OCD (P1, P8) found familiar triggers less intense. For instance, P1 was not distressed by food residue if she knew who was responsible for it.
\textit{``Right now, in my kitchen, there's a few dirty dishes in the sink, and there is a pot on the stove, and a box of pasta that needs to be put away, but I know that my roommate made Pasta last night, and so I'm not at all stressed. Because I know exactly where everything came from.''} 
Similarly, P8 was more concerned about germs brought into his apartment by strangers than by his own family, whose hygiene he trusted.

\textbf{\textit{Internal Factors Affecting Trigger Intensity.}}
In addition to trigger properties, participants with OCD highlighted several internal factors that influenced trigger intensity. Six participants (P1-4, P6, P7) further pointed out that their OCD triggers intensified when experiencing \textit{mental stress}. It can be from schoolwork (P1, P2, P4), being in unfamiliar environments (P1) and personal life issues (P4). Additionally, poor self-care, such as sleep deprivation and physical inactivity, also exacerbated trigger intensity (P1, P2). Interestingly, unlike mental stress, \textit{physical stress} could sometimes help reduce trigger intensity (P7, P8). As P8 explained, \textit{``if I'm pretty tired, my brain can't put so much energy towards the OCD.''} Similarly, trigger intensity can be reduced when \textit{short-term memory} is fully occupied by other tasks (P2, P6), e.g., when working (P6). 

\mybox{\textit{Inspiration}:} Trigger intensity is affected by not only trigger itself, but also properties of triggers whose variation can result in significantly different anxiety levels. As a result, trigger recognition for just-in-time intervention should not only consider object-level detection but also the recognition of certain properties to more accurately predict users' distress, thus providing suitable intervention.

\subsection{How Triggers Shape Compulsions}
\label{compulsion types}
Besides triggers, compulsion is another important indicator to invoke just-in-time interventions. Timely detection of compulsive behaviors in daily life is crucial for delivering effective interventions. We thus investigated the diverse compulsive behaviors and their relationships to triggers to inspire compulsion detection. 
\subsubsection{Distinguishing Compulsive Behaviors}
Compulsive behavior is harder to identify than triggers since compulsions often manifest in ways people with OCD might not even realize, such as avoidance (e.g., avoiding contact with objects deemed dirty by people with contamination OCD) and mental compulsion (e.g., rumination) (T1-T3, T5-T7). Therefore, in addition to identifying the typical compulsions (e.g., excessive hand washing) using standardized checklist (e.g., Y-BOCS \cite{goodman1989yale}), professionals would also ask clients about the presence of covert compulsions directly, such as avoidance and mental compulsions (T1-T3, T5-T7). In general, our professionals outlined three criteria to help distinguish compulsive behaviors from regular behaviors and assess their impact on daily life:
\begin{itemize}
    \item \textit{Functional Impairment} (T1): Evaluating how a behavior limits essential activities, such as avoiding social interactions due to intense mental compulsions (e.g., avoiding going outside).
    \item \textit{Behavioral Rationale} (T5-T7): Determining whether the behavior serves a specific purpose, or if the goal is merely to reduce anxiety (e.g., frequently seeking reassurance from others).
    \item \textit{Common Sense} (T7): Using common sense to assess whether a behavior is compulsion by considering what others would do to achieve the same goal (e.g., excessive hand washing).
\end{itemize}

\mybox{\textit{Inspiration}:} These compulsion identification strategies from professionals could potentially be adapted to support compulsive behavior recognition in real life, for example,  with the help of machine learning models that can distinguish common sense from irregular behaviors \cite{aggarwal2021explanations}.


\subsubsection{Connecting Compulsions to Triggers}
Similar to OCD triggers, our OCD participants experienced diverse compulsive behaviors. Unlike prior OCD literature that focused on recognizing compulsive behaviors in itself (e.g., washing hands) \cite{starcevic2011functions, goodman1989yale}, we found that our participants constantly associated their compulsive behaviors with triggers. Understanding the rationale behind compulsions, particularly how individuals respond to specific triggers, is crucial in interpreting compulsive behaviors, which can potentially help enhance compulsion detection and facilitate intervention delivery at the right timing. We thus categorize compulsions into five categories based on their relations to triggers:

\textbf{\textit{Direct Trigger Interaction.}} 
Direct trigger interactions are physical behaviors that engage with the trigger to temporarily resolve the distress caused by the trigger. Repeatedly examining the trigger is a common compulsion for participants with responsibility OCD (e.g., checking or jiggling the door locks, P8, P9), and contamination OCD (e.g., checking the color of meat being cooked to ensure doneness, P3). Ordering and organizing is another direct interaction with triggers to precisely control the position, orientation, or order of the triggers to avoid accidents (e.g., keep appliances unplugged when not using, P7, P8), reduce harmful thoughts (e.g., pointing the knife to herself to protect others, P2), and keep life organized and ``just right'' (e.g., having to park his car in the same spot, P10). 
Direct trigger interactions can also happen when the OCD trigger is a person's body part. For example, P4 would pull her eyelash or hair so both sides look even whenever she examines her look in the mirror.
However, while temporally relieving mental distress, such compulsion could bring long-term mental health issues, \textit{``[Hair and eyelash] takes the longest to grow back and it really impacts my confidence and stuff, because it's noticeable.''} (P4)


\textbf{\textit{Trigger Avoidance.}}
Trigger avoidance is another type of compulsions that minimize the interaction with triggers (P1, P3, P6, P8), 
We found that the extent of avoidance was a spectrum, ranging from partial avoidance, such as rolling their hands into sleeves (P3, P8), or using as few fingers as possible (P6) when touching door knobs, to complete avoidance, such as avoiding eye contact with the trigger (P2). 
Trigger avoidance can lead to severe functional impairment, threatening OCD people's quality of life. P8 expressed his struggle to engage in essential daily activities due to his contamination avoidance, 
\textit{``If there are certain activities that I think are dangerous that's not safe or something bad could happen. It's gonna get in my head... The biggest challenges are just refusing to engage in a lot of activities to avoid... 
And I'm not gonna be able to fix the issue the way I want. And I'm gonna be stuck somewhere where I can't do anything about it. 
''}

\textbf{\textit{Consequence Mitigation.}} 
Consequence mitigation is a compulsive behavior that addresses the speculated aftermath after interacting with an object trigger when it is inevitable. For participants with contamination OCD, changing/cleaning clothes when coming back home and washing body parts after contacting something `dirty' are common compulsions. Additionally, P3 mentioned self-body scanning for potential symptoms when encountering someone appearing sick. For participants with responsibility OCD, repeated checking can also be consequence-oriented. For example, participants who worried about hitting pedestrians while driving would check for marks on the car later (P7) or look at rear mirror to confirm no person was run over (P9). P7 described this lengthy process, \textit{``It used to take me like 10 minutes to check all of my car because I would get out and I would check like the doors and the windows, and I would check the hood of the car and the back of the car for scratches.''}

\textbf{\textit{Assisted Compulsions.}} 
In addition to performing compulsions independently, some participants sought accommodations from others (P1, P3, P6-8, P10) to help them complete compulsions. For instance, P10 disclosed their OCD to gain understanding when their symptoms affected others. 
He described the struggle of resisting a compulsion and how he ultimately requested someone to move their car from his preferred parking spot,
\textit{``I went and knocked on the person's door and told them to move their car, as unreasonable as that sounds. At this moment, I physically couldn't stop myself, I had to. It was a compulsion I could do nothing about... 
I'm sitting here sweating, crying... you know, almost in physical pain, over the fact that my car is not in its `correct' spot. And then I just said, `Okay, I have to go talk to this person.' 
''}

Some participants also proactively asked (P1, P8) or passively waited (P3, P7) for help from other people to share the burden of performing compulsions. For example, P8 asked his friends to reassure him that the faucet was off instead of checking it repeatedly, since he trusted them more than himself. P3 would feel relieved when someone held doors for her to avoid contamination. 

\textbf{\textit{Mental Compulsions.}}
In addition to physical compulsive behaviors, seven participants reported experiencing mental compulsions (P2-5, P7, P9, P10) where they attempted to neutralize the influence of triggers internally. For participants with just-right OCD, a common mental compulsion was internal counting, for instance, internally counting delayed time when their schedule was disrupted (P10). Other mental compulsions included self-questioning rumination, such as convincing oneself of being a good person despite having intrusive thoughts (P2, P5), catastrophizing about consequences of disrupted routines (P10), and self-reassurance to suppress negative thoughts (P7).

\textbf{\textit{Effect of Environment Familiarity on Compulsions.}}
Interestingly, five participants argued that the length and frequency of their compulsions was affected by their familiarity with the environment. P6 had more compulsive behaviors when around closed people than strangers, and four other participants (P2, P4, P8, P10) experienced fewer compulsions in public space. P4 indicated that she did not want others to see her compulsions, while P6 and P10 `gave up' on compulsions in public knowing that they have no control over the environment. 

\mybox{\textit{Inspiration}:} Connecting triggers and context information to compulsions can potentially facilitate compulsion recognition by tracking not only user behaviors but also how specific triggers shape the compulsive reaction of the user. As such, a user's compulsion patterns can be accurately characterized, allowing for more contextualized and personalized intervention design.

\subsection{Managing OCD at Different Stages}
\label{management strategies}
To manage their OCD symptoms, our OCD participants adopted a range of strategies learned from OCD therapy (P1–5, P7, P9, P10), support groups (P5), online educational resources (P8), and personal experiences (P6, P10). These strategies were applied at different stages of an OCD episode—from the onset of obsessive thoughts to the execution of compulsions. By integrating perspectives from both people with OCD and professionals, we observed substantial overlap in strategy use, but also noted some inconsistencies in how these strategies were applied or recommended. We summarize these coping strategies to inform the \textit{\textbf{feedback design of just-in-time interventions}} for OCD self-management technologies. 

\subsubsection{Accepting \& Challenging Obsessive Thoughts}
OCD participants employed different mental strategies to address their obsessions, adapted from ACT and cognitive therapy (see Section \ref{def ocd}).

Three participants with OCD and nearly all professionals (except T4) used or recommended ACT-based strategies to help shift focus from obsessions to value-driven actions. These strategies included: 
(1) \textit{Thought Diffusion (P3, P5, T1, T5)}---recognizing and naming the intrusive thought to create psychological distance from the thought; 
(2) \textit{Embracing Uncertainty (P2, P3, T5-T7)}---using self-talk like \textit{``maybe, maybe not''} to accept uncertain outcomes; 
(3) \textit{Refocusing on Values (P5, T1-T3, T5)}---identifying values (i.e., important aspects of life) and committing to behaviors aligned with those values.

While some OCD participants benefited from accepting obsessive thoughts, 
others managed obsessions by actively challenging and re-evaluating them to \textit{reinforce positive thinking} (P1, P7), a technique also endorsed by therapists (T3, T5). For example, P7 addressed her intrusive fears of unconsciously hitting a pedestrian while driving by evaluating the evidence:
\textit{```okay, what's the likelihood that this (hitting pedestrians) actually happened? And you know, wouldn't someone else have noticed if this had happened? Wouldn't I have heard, or seen, or touched or smelt things if this had happened?' So it address like the thought pattern.''}

\mybox{\textit{Inspiration}:} Leveraging the self-reflective nature of these mental strategies, we can design conversation-based persuasive intervention \cite{orji2018persuasive} to guide people with OCD to adopt beneficial behaviors and thoughts when obsessions occur, such as intelligent agents that help users navigate their thought processes and address cognitive pitfalls.

\subsubsection{Compulsion Prevention via Real Life Exposure Practices}
When obsessive thoughts were too intense to address or dismiss, participants with OCD resorted to strategies learned from ERP (P1-3, P5, P7, P10) to effectively train their ability to tolerate the anxiety without performing compulsions.

\textbf{\textit{Planned vs. Natural Exposure Practice.}} 
To improve symptom management ability in a controlled environment, participants would do \textit{planned exposure practice} as instructed by their therapists (P1-3, P5, P7, P10). In a typical planned exposure practice, people with OCD expose themselves to a self-created triggering environment starting from triggers with moderate intensity on the trigger hierarchy at a scheduled time, and resist compulsions.
P3 articulated her exposure practice targeting the fear of touching batteries,
    \textit{``It (exposure practice) was holding a battery in my hand, rubbing the battery on my arms, putting it in my pocket, rubbing it on my clothes, my phone, on my bag... So I, essentially contaminating all these different things [with batteries] and then, not being able to wash my hands or do anything to decontaminate the battery acid... and having to sit with worrying that everything was contaminated.''} 
While tolerating the mental distress, people with OCD were instructed to monitor their anxiety level change (e.g., SUDS level) (T2-T6) during the exposure practice. In addition to self rating, when possible, professionals detected clients' anxiety level also by observing their physical behaviors (T3, T7), such as fidgeting, head shaking, tight chest, staring off the trigger, or sometimes an outward expression of distress. Typically an exposure practice concludes when their anxiety level decreases (T2, T4, T7). 

Although the flexibility of adjusting trigger intensity allowed participants to gradually build up their tolerance to triggers, most triggers were unplanned in real life, and some triggers that induce intrusive thoughts could not be anticipated (P2, T1, T7). 
Therefore, when confronted with unplanned triggers, participants would seize the opportunity to do a \textit{natural exposure practice}. 
Due to the spontaneity of unplanned triggers and unpredictable trigger intensity, natural exposure practices were perceived to be more difficult (P1-P3, P7, P10). 

Despite the effectiveness of both planned and natural exposure practices, OCD participants found them challenging especially at the beginning (P2) and tended to avoid exposures when alone (P1). They had to force themselves to complete the practices, since compulsions were too difficult to resist (P1-P3, P10). As P10 described, during an exposure practice where the trigger was overwhelming,  \textit{``it was mentally painful to try and stop (the compulsion).''}

\mybox{\textit{Inspiration}:} These findings highlighted the importance of providing timely and context-aware guidance to help users engage with the exposure rather than avoid it, especially at the beginning, when the user is still building confidence in managing their symptoms.

\textbf{\textit{Strategies for Resisting Compulsions.}} 
To better support compulsion prevention, OCD participants reported different mental and physical strategies to suppress the urge of doing compulsions. One common strategy is to \textbf{re-evaluate the necessity of compulsions} mentally (P2-4, P6). By \textit{reflecting on past OCD episodes}, participants reminded themselves that nothing bad happened when compulsions were not done (P2) and that compulsions were never working (P4, T1, T7). 
Some \textit{weighed the harm compulsions caused against the relief they brought} (P3, P4). For instance, frequent hand washing killed germs but caused dry, irritated skin (P3).
In addition to mental persuasion, participants also employed physical strategies to \textbf{make compulsions more difficult to act on}, e.g., P4 creatively put Vaseline on her eyelash and wore glasses to prevent herself touching her eyelash. 

Completely resisting compulsions overnight can be overwhelming. Instead, gradually tapering them—by \textbf{allowing milder forms of compulsive behavior}—can support long-term compulsion reduction. Both OCD participants and professionals in our study employed such strategies. Participants with responsibility OCD used \textit{reassurance logging} to gradually reduce the intensity and frequency of their compulsions over time (P8, P9). For example, they recorded videos or took photos of their checking behaviors to ease anxiety and avoid the urge to re-check later. As P8 explained:
\textit{``If I know I'm gonna be gone for a while. I will take a video where I'll video my entire apartment, and then I'll also turn the camera around and say, `I have checked the faucet this many times; all plugs are unplugged...' so it's almost like telling myself through a video in the moment that I'm good to go. And then if I ever get anxious, I can look back at the video.''} 
Similarly, professionals allowed clients to engage in reduced compulsions when full exposure practice felt too distressing. These included compulsive behaviors with shorter duration (e.g., washing hands for a shorter time, T3, T7) and lower frequency (e.g., checking less often, T2).

\mybox{\textit{Inspiration}:} In summary, participants effectively reduced the urge for compulsions by drawing on past behaviors and memories of previous OCD episodes. Inspired by these strategies, symptom and behavior logging could be leveraged to design just-in-time intervention methods that increase the cognitive and physical burden of engaging in compulsions.
Tapering strategy can also be translated into just-in-time interventions by prompting users with milder alternatives to their compulsions, helping them gradually build tolerance to anxiety.


\textbf{\textit{What is Missing in Exposure Practice?}}
While all participants were confident about the effectiveness of exposure practices, many acknowledged their limitations in addressing mental compulsions (P2, P5, T1, T2). These covert compulsions, such as self-reassurance or repeated mantras, often persisted even when physical compulsions were reduced.
T1 highlighted this challenge, explaining that some clients might compensate for their behavioral change with mental compulsions,
\textit{`` They (clients) are doing mental compulsions to sort of cancel out the fact that they didn't do the physical compulsion, like, okay, I didn't [check] that doorknob. Okay, I'm making my therapist happy, but what I'm doing in my mind is, I'm going over my mantra 15 times, because that makes me safe.''} To compensate this limitation, some professionals identified clients' engagement of mental compulsion alongside their physical compulsive behaviors. For example, T2 described asking clients to rate their mental effort during exposure practices to capture these hidden patterns.

\mybox{\textit{Inspiration}:} This highlights the importance of detecting and monitoring mental compulsions in the design of just-in-time intervention technologies to reveal the full picture of a user’s compulsions.


\subsubsection{Distraction \& Relaxation as Safety Net}
When their OCD became so overwhelming that all above strategies failed, OCD participants (P1, P3-P5, P8, P10) turned to \textit{self-distraction} as a temporary escape---shifting their attention to casual activities such as watching TV (P3), listening to music (P8), calling a friend (P1), or going out for a walk (P5). Similarly, \textit{relaxation techniques} were used to ease anxiety when breaking the obsession–compulsion cycle felt impossible (P1, P3, P6, P7, P10). Such strategies included \textit{breathing techniques}, \textit{muscle relaxation}, and \textit{grounding exercises}. 

While professionals supported using these methods to manage intense anxiety (T1, T2), they cautioned that such strategies must be used carefully. They noted the importance of helping clients learn to tolerate anxiety rather than eliminate it (T1, T2, T5, T6). Without careful guidance, overreliance on distraction or relaxation can itself become a form of compulsion (T1, T2, T5).
As a result, T7 chose not to teach relaxation techniques at all because they might backfire if they don’t work when clients need them most, \textit{``At one point, [the client] may be really distressed. And they've been practicing these breathing techniques that are not going to work for you, then it's even more distressing, because the only thing you know how to do [with your OCD] is not gonna work.''}

\mybox{\textit{Inspiration}:} Relaxation strategies can be beneficial to people with OCD only when used correctly. It is important for just-in-time technology to involve proper guidance to support and prevent relaxation at the right timing.





\subsubsection{External Mental Health Support}

Psychotherapy is a key resource for learning OCD coping strategies \cite{o2015cognitive}. While most participants reported positive experiences, some felt lost and discouraged in therapy not specialized for OCD (e.g., generic talking therapy for P6). Nine out of 10 participants also experienced challenges accessing OCD therapy due to long wait time (P1, P5–P7, P9, P10), high costs (P1, P3–P5, P7–P9), and stigma around OCD (P6, P7), underscoring the need for supplementary forms of support.

Beyond formal therapy treatment, most participants (P1–8, P10) received support from family and friends to help manage their symptoms. Some had close contacts who understood their condition and followed therapeutic guidance---helping monitor symptoms (P4) and encouraging them to challenge their OCD independently (P1–3, P7).

Others, however, had family members or friends who misunderstood OCD and unintentionally reinforced symptoms by providing accommodations contrary to therapists' suggestions. For example, P6’s mother washed her hands every time after using elevators to accommodate P6’s contamination-related fears. As several professionals emphasized (T1–5), \textit{psychoeducation} on how therapy works and the expected outcome of OCD therapy for family members is critical, so they can respond to their loved ones in ways that support, not hinder, treatment progress (T2–4).

\mybox{\textit{Inspiration}:} Psychoeducation could be integrated into OCD self-management technologies to improve awareness and collaboration between people with OCD and their support system.


\subsection{Opportunities and Gaps in OCD Self-Management Technologies}
In this section, we explore participants’ experiences and attitudes toward existing technologies, and identified key challenges in OCD self-management despite the availability of technological support. Based on this, we discuss opportunities to better support OCD self-management.

\subsubsection{Technology Recommendation for OCD Self-Management}
Surprisingly, none of our OCD participants reported using technology specifically designed for mental health support. In contrast, OCD professionals recommended a range of tools from symptom tracking to compulsion prevention to help clients manage their OCD outside of therpy (T2-5, T7). Dedicated OCD symptom tracking apps, such as \textit{OCDfeat} \cite{ocdfeat}, were suggested to easily record compulsions and exposure practices (T2). They also recommended meditation apps, such as \textit{Calm} \cite{calm}, for clients to practice mindfulness (T5) and listen to the bilateral stimulation playlist to calm the nervous system (T4) \cite{kohl2014deep}. These tools are intended for consistent out-of-session use to monitor symptoms and reinforce skills learned from the therapy.

Professionals also recommended apps for in-situ use in response to OCD episodes upon triggered. To reduce reassurance seeking behavior using specific apps (e.g., web browser), T3 suggested app redirection tool (e.g., \textit{ScreenZen} \cite{screenzen}) to limit access to those apps on clients' phones.
Interestingly, he also recommended text replacement feature to clients to prevent them from searching specific keywords for obsession-related reassurance, \textit{``
I'll have people create shortcuts where... if they're going to Google `heart attack,' it actually will try to replace `heart [attack]' with like `puppies' or `cute cat videos,' or something like that.''} 

Although most participants prefer learning and recommending new technology to their clients, some were reluctant to introduce new tools in the fear of developing reliance on them (T1, T5). T1 expressed her concerns, 
    \textit{``You want to encourage people to instead learn how to cope with things by themselves... right? That's what we ultimately want. So maybe you could have [the technology] where it was only like the first month or the first 3 months, or whatever, and then you have a gradual step down, that might work... because [relying on technology] is counter-therapeutic. We want people to be able to help themselves.''}
    
When talking about self-administered ERP via mobile apps (e.g., NOCD \cite{nocd}), professionals recognized its merit in supporting users with low severity (T1-4). However, they emphasized the importance of psychoeducation in OCD and acquiring adequate symptom management ability via in-person sessions before using such self-help tools (T1, T3, T4, T6, T7).

\mybox{\textit{Inspiration}:} To prevent overreliance on technology becoming a new compulsion, just-in-time intervention should adopt a wind-down strategy to gradually help users be independent on symptom management.

\subsubsection{Inconvenience \& Privacy Concerns in Symptom Tracking}
Daily symptom tracking helps people with OCD build self-awareness around their behaviors and thoughts (P1, P4, P5, P8). It also allows them to observe progress and recognize achievements (e.g., spending less time doing compulsions) over time, which can foster a sense of accomplishment, increase confidence, and strengthen motivation to continue practicing coping strategies (T2–T4, T6). Participants typically logged daily triggers (P4, P8), their responses to those triggers (P1, P4), and distress level during exposure practices (P1, P5) either physically or digitally (e.g., Google Doc). 

However, maintaining a consistent tracking habit was challenging (P1, P2, P7, P8, T3, T6). OCD participants found traditional journaling methods superficial, as it failed to provide meaningful insights into participants' symptom patterns or suggest optimal management strategies (P1, P8). Professionals, on the other hand, attributed low engagement to forgetfulness due to other daily responsibilities (T3, T6). To address this, professionals sometimes had clients fill out symptom tracking forms and set up phone reminders to ensure compliance (T3).

Privacy concerns also hindered the use of symptom tracking tools (P7, T2). Participants were especially cautious about recording sensitive thoughts or exposure practices without clear information on how their data would be stored and protected. P7, for example, expressed worry about the emotional risk of recording vulnerable moments:
\textit{``I would be most concerned about privacy issues, because it's very vulnerable for people... to do all of that exposure [practices] in a place where their data is [at stake]. So if sessions are recorded... or their speech or their notes are recorded, making sure all of that is well protected...
''} Similarly, T2 noted that clients whose OCD symptoms involved taboo thoughts often avoided logging due to fears about data privacy and access, as she explained, \textit{``
They (clients) do worry about putting that (taboo thoughts), even writing it down on paper... so they need a lot of reassurance about where that data is being stored and who has access to it.''}

\mybox{\textit{Inspiration}:} Symptom tracking is a crucial first step for enabling effective just-in-time interventions. These findings highlight the need for a simple, transparent tool that can automate tracking while respecting and protecting users' privacy.

\subsubsection{OCD-Induced Social Barriers}
In addition to challenges with self-management, participants with OCD also described how OCD created difficulties in their social lives. Due to the heterogeneity of participants' OCD symptoms, understanding the nuanced experience of OCD could be difficult, even with OCD professionals (P6, P10). P10 had to downplay his symptoms to gain empathy from his therapist. 
Outside of therapy, participants often felt their compulsions interrupted social interactions (P2, P7, P8, P9). For instance, P9 found it embarrassing to go back to check the parking brake of his car during group activities. 
Moreover, exposure practices in real life can also bring social barriers since practicing in public spaces or when other people are around could be socially awkward (P1, P7). 

\mybox{\textit{Inspiration}:} More support is needed across social, clinical, and technological contexts to foster greater awareness of OCD and ensure that people with OCD have access to supportive environments and resources.

\section{Discussion}
As an important aspect of quality of life, self-management of OCD presents significant challenges and the needs for technological support has long been underexplored. Our research fill this gap by thoroughly investigating the perspectives of both people with OCD and OCD professionals. Our findings centered on the lived experiences of people with OCD and supplemented and contrasted with the professionals' views on treatment strategies and technology recommendations.   
As a result, we identified the gaps and opportunities in OCD symptom self-management. 

First, while natural exposure practice was encouraged due to their abundant presence in daily life, these practices were often difficult to complete and tended to be avoided \cite{seibell2014management, eddy2004multidimensional}. Besides the difficulty of facing the fear, identifying mental compulsions could be challenging and required adequate insights. The absence of physical compulsion does not guarantee the absence of mental compulsions despite going through ERP.
Second, some experience-based strategies adopted by people with OCD (e.g., relaxation techniques) were 
suggested to be used cautiously by the professionals due to the potential risk of developing reliance.
Third, both OCD participants and therapists recognized the importance of symptom and activity tracking, however, traditional tracking tools present limitations and raise privacy concerns due to vulnerability and shame associated with OCD, resulting in low technology adoption.
Finally, people with OCD faced persisting challenges in OCD self-management especially in social scenarios despite receiving therapy, highlighting the need for additional support to facilitate smooth social interactions.

These gaps emphasized the needs for self-management technologies beyond formal therapy sessions. Based on the diverse OCD triggers, compulsive behaviors, and symptom management strategies identified in our research, we discuss the \textit{input (e.g., trigger and compulsion recognition)} and \textit{output (i.e., feedback design)} of future just-in-time self-management technologies for OCD. Furthermore, given OCD people's challenges, we discuss how technology can address OCD-specific privacy issues and opportunities to support OCD people's social needs.


\subsection{Enhancing Symptom Recognition \& Tracking with User Input} 
As an important step of establishing symptom self-awareness for symptom self-management, our study underscored a need for supporting personalized, automatic, and fine-grained tracking of OCD symptoms and exposure practices. 

\subsubsection{Supporting Personalized \& Automatic OCD Tracking}
We found that OCD triggers are highly individualized, taking diverse forms (e.g., objects, contexts, sounds, thoughts), and that their properties significantly influence trigger intensity. While symptom tracking is essential for improving symptom awareness and motivation, it is difficult to maintain without therapist supervision.
Therefore, future technology should incorporate OCD users' input, allowing specification of triggers to support personalized and automatic trigger detection. Similarly, the system can prompt for the user's compulsive behaviors based on different types of compulsions and criteria proposed by OCD therapists (Section \ref{compulsion types}). 
Recent advances in AI have made it possible to adapt pre-trained models to specific domains through few-shot learning \cite{parnami2022learning, zhang2023prompt}. By leveraging just a few examples provided by users, these AI models can be personalized to effectively detect and track OCD symptoms within each user's unique context.
Given the varied form of triggers and compulsions (e.g., direct trigger interaction, consequence mitigation), multi-modal AI models \cite{xu2024survey, xie2024large, bordes2024introduction} can be used together with wearable devices (e.g., egocentric camera and microphone \cite{liu2024human}) to identify and track triggers and compulsions beyond the scope of vision-based methods. Despite the merit of AI-based symptom tracking automation, false positive prediction results can potentially exacerbate OCD symptoms, since people with OCD often have low confidence in their memories \cite{radomsky2014more}. To minimize this risk, prediction models should communicate potential errors and involve users in assessing the decision-making process using explainable AI techniques \cite{hossain2023explainable}.

\subsubsection{Fine-Grained Symptom Formulation} 
Besides highly personalized triggers, our findings suggest the importance of considering different properties of triggers to enable fine-grained exposure hierarchy building and symptom tracking. In complement to personalized trigger specification, future tracking technology should prompt the user to reflect on how trigger intensity is affected by different properties of that trigger, measured with SUDS or urge to perform compulsion (Section \ref{props of triggers}). As such, an exposure hierarchy with fine-grained and structured trigger properties can be built. 
During tracking, the system can log fine-grained details of triggers and situate them in the hierarchy as well as identify physical compulsive behaviors, to better support the tracking of exposure practices. 

\begin{figure*}[t]
    \centering
    \includegraphics[width=0.8\textwidth]{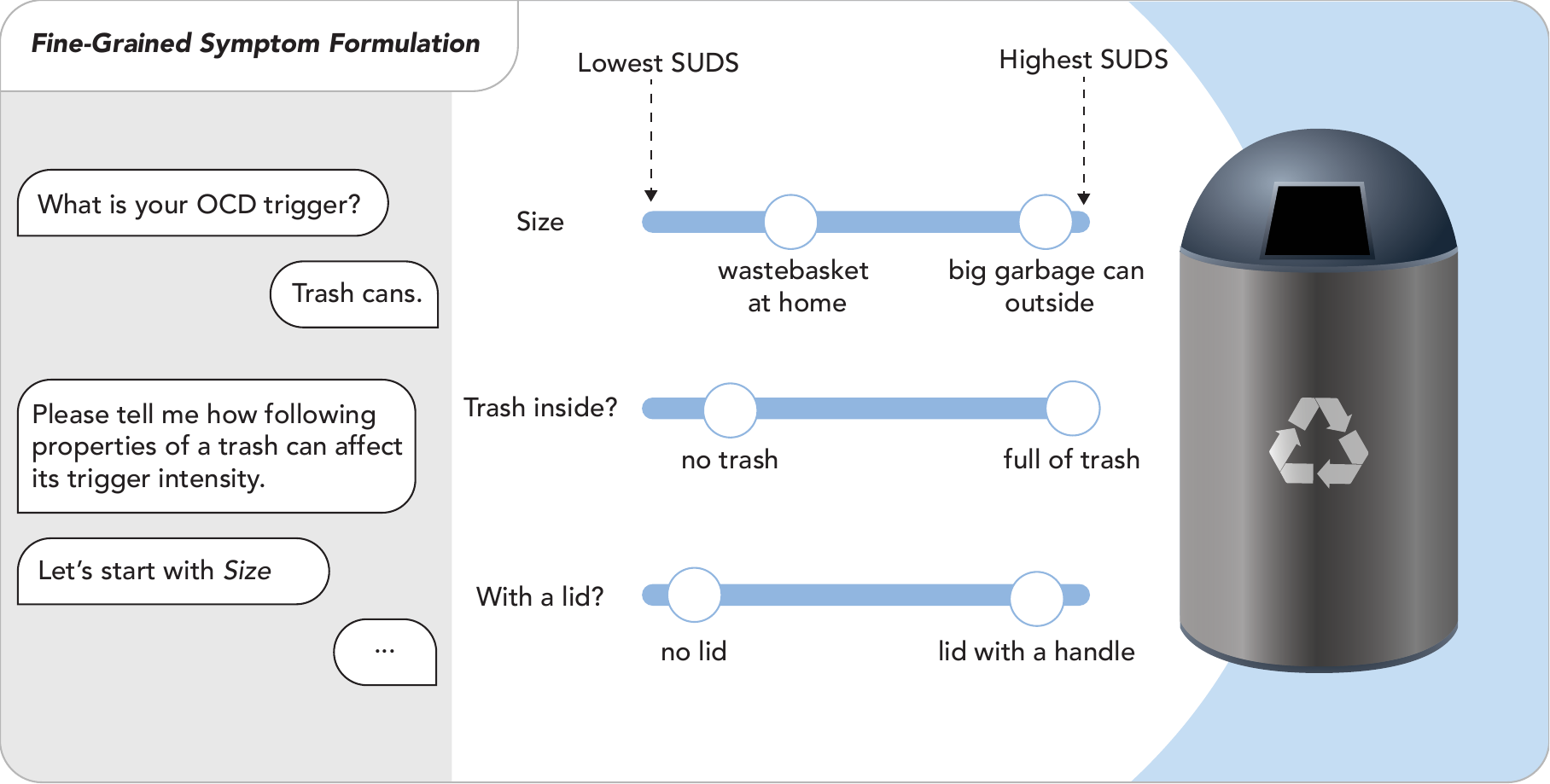}
    \caption{Fine-Grained Symptom Formulation: an example of the user specifying triggers and trigger properties to build fine-grained exposure hierarchy.} 
    \Description{The figure mainly contains three part left to right. On the left, it shows a conversation interface between a user with OCD and an AI agent. The agent asks the user ‘What is your OCD trigger?’ the user answers ‘Trash cans.’ Then the agent says ‘Please tell me how following properties of a knife affect its trigger intensity. Let's start with Size.’ The user’s speech bubble shows ‘...’ In the middle, there are three sliders labeled with different properties. All three sliders are of same length with the left end representing ‘Lowest SUDS’, and right end representing ‘Highest SUDS’. There are two knobs on the ‘size’ slider saying "wastebasket at home" and "big garbage can outside" located at the middle-left and the right end of the slider, respectively. There are two knobs on the ‘Trash inside?’ slider saying "no trash" (located near the left end of the slider) and "full of trash" (located on the right end). There are two knobs on the ‘With a lid?’ slider saying "no lid" (located near the left end of the slider) and "lid with a handle" (located near the right end). On the right of the figure, it shows an illustration of a silver trash can with a recycling symbol on it.
}
    \label{fig:implication}
\end{figure*}

Figure \ref{fig:implication} shows a usage scenario that can be achieved with proposed system: the user identified ``trash can'' as a trigger of their contamination OCD, they then are prompted to answer how each property can affect the trigger intensity, for example, a trash can can be more triggering when the trash can is overflowing, so they assign ``full of trash'' with the highest intensity level, and ``no trash'' with a moderate intensity level. They are also prompted to describe their compulsion---``avoiding using the trash can.'' The system will then detect the physical triggers and compulsions in real-time from the camera feed.

Mental compulsions are often difficult to recognize but critical to address. To support their detection, future technologies could incorporate physiological sensing, such as EEG and ECG, to assess mental workload \cite{fan2020assessment, wang2023research}, inspired by strategies used by professionals (T2). Prior research has also shown promising results in detecting mental rumination using self-report measures combined with wearable sensor data, like EMG \cite{nalborczyk2017orofacial, arnold2025potential}. By capturing a user’s full compulsion pattern and integrating contextual information (e.g., triggers, environment), the system can generate more tailored and effective intervention strategies.




Despite the potential,
prompting users to report symptoms itself can be triggering. Our suggested input methods may require a certain level of symptom awareness and readiness for symptom management, which should be assessed before prompting users' symptoms. Addressing this challenge is essential for creating safe and effective self-management tools. Future research should explore symptom reporting methods that minimize discomfort while ensuring honesty.

\subsection{Just-in-Time Feedback Design for OCD Self-Management} 
Consistent with prior literature \cite{lind2013technological,hezel2019exposure,ojalehto2020adherence}, exposure practices can be difficult to follow through outside of therapy sessions, due to the existence of fearful triggers in diverse context. 
Triggers can appear anywhere anytime and provoke users' OCD spontaneously. As such, the burden of dealing with unpredictable triggers and resisting compulsions all fall onto the users with OCD. This can potentially result in severe consequences. We thus propose the following technologies that provide just-in-time intervention during high intensity exposures in people's everyday life.



\subsubsection{Augmented Reality-Based Compulsion Prevention Support} 
Inspired by OCD participants' strategies and OCD therapists' technology recommendation, future intervention technology should consider using \textit{restrictive} interactions to make compulsions more difficult to perform. 
Inspired by prior work on smartphone overuse prevention \cite{orzikulova2024time2stop, lu2024interactout, xu2022typeout}, future intervention technology can adopt similar restrictive interactions to help reducing compulsions in digital space, e.g., seeking reassurance online.
For compulsive behaviors that happen in physical world, potential technology can explore the design of augmented reality (AR) -based interaction to suppress the urge for compulsions. Researchers has explored the usage of visual barriers in AR and VR (virtual reality) systems to ensure user safety \cite{hoang2022virtual, sousa2019safe}. Such design can be adapted for compulsion prevention. 
Since most overt compulsions involve visual examining (e.g., repeated checking), adequately disturbing the user's vision by placing visual barriers on objects being examined may effectively interrupt physical compulsions. For instance, rendering visual blockage on stovetop when the user repeatedly checking burner knobs. 


For trigger avoidance type of compulsions, such as avoiding eye contact with people that induce intrusive thoughts (P2), the system can render visual augmentations that encourages interactions with the trigger. For instance, when avoidance of triggers is detected, the system can render a gray-scale filter on everything the user sees except the triggering person to encourage eye contact, and turn off the filter when eye contact is maintained for a preset length of time as an exposure practice. 

Despite the benefits of restrictive interactions in compulsion prevention, overly restrictive interactions might lead to low user compliance and even backfire \cite{lu2024interactout}. Future work should explore the optimal balance between effectiveness and restrictiveness of such technologies in OCD context.

\subsubsection{LLM-based Cognitive Support}
In addition to physical methods, mental strategies inspired by ACT and cognitive therapy are essential for helping individuals reshape their attitudes toward obsessions. LLMs can be leveraged to offer personalized suggestions based on these strategies when mental distress or compulsions are detected. For instance, the technology can guide the user to embrace uncertainty following a triggering event. 
Moreover, \textit{value-based} intervention techniques \cite{khan2011use, xu2022typeout} can serve as a `wake up call' to remind the user of committing to their values---for instance, displaying a photo of a loved one on the user’s phone to interrupt compulsive hand-washing, if the user has identified ``spending time with family'' as a core value.
Alternatively, we can design intervention based on thought reframing---guiding the user via conversations to challenge their obsessive thoughts and reassess the likelihood of feared outcomes \cite{sharma2023cognitive, yule2017cognitive}. 

However, generating effective, context-aware responses requires carefully crafted prompts, which can be difficult without domain-specific expertise \cite{priyadarshana2024prompt}. Although prior work has explored LLM-based support for general CBT \cite{kian2024can, huang2024empowerment}, no existing study has focused on adapting these techniques for OCD-specific contexts. Future work should explore not only prompt engineering for pre-trained LLMs but other techniques such as retrieval-augmented generation \cite{lewis2020retrieval, guo2024soullmate} to integrate OCD-specific knowledge and user history for more targeted and effective intervention.


\subsection{Addressing Privacy Needs for People with OCD} 
Privacy concerns have been discussed in prior literature about personal information sensing and tracking \cite{harari2020process, torous2018clinical}. To preserve users' privacy, the tracking technology should explicitly inform them about what kind of data are being collected, what kind of data are inferred from collected data, and who has access to collected data \cite{harari2020process}. When it comes to AI models, on-device inference is preferred to minimize the risk of privacy invasion \cite{lee2019device}. However, due to technology limitation, many AI services require sending user data to remote servers for inference \cite{chatgpt}. In the context of OCD, researchers should pay attention to sensitive information that can exacerbates OCD users' mental health conditions. Users should be prompted to identify the type of sensitive data, such as audio or video involving fear, and taboo thoughts (e.g., sexual and violent thoughts). The system should then actively detect sensitive information in collected data and obfuscate them, such as blurring out sensitive elements in videos and photos \cite{zhang2024designing, tseng2024biv}, and reframing sensitive text to less disturbing alternatives \cite{chatgpt} before sending them out for inference. The collection of biometric data (e.g., physiological data for mental state inference) can also pose privacy concerns due to potential unauthorized use of those data to infer personal identity \cite{marciano2019reframing}. Future OCD technology should adopt privacy-enhancing approaches such as feature transformation \cite{melzi2024overview, nandakumar2015biometric} to lower the risk.

\subsection{Reducing Frictions in Social Interactions}
Our study suggest that OCD poses challenge to smooth social interactions due to the need for compulsions. Prior work also revealed that people with OCD experienced more interpersonal problems primarily due to certain worries and rumination than people without OCD \cite{solem2015interpersonal}. 
Due to the low public awareness of OCD, people with OCD usually need to endure the awkwardness caused by their symptoms during social interactions. Disclosing one's OCD status can earn accommodation or understanding from people (P10). However, not all people with OCD are willing to disclose their mental health status, particularly when their OCD involves taboo thoughts (e.g., P2's sexual OCD). For those who are still building symptom management skills, additional support 
is needed to reduce frictions caused by OCD in everyday social interactions.

With current technology, researchers can explore the use of motivational interviewing via conversational agent \cite{miller2012motivational, oewel2024approaches} to encourage the user to keep engaged with social interactions. 
Technology that delivers unobtrusive intervention for emotion regulation can also be used to address the mental distress in such situation
\cite{costa2016emotioncheck, costa2019boostmeup, kelling2016good}. 
That said, technology can provide limited assistance to satisfy OCD people's social needs without improved public awareness of OCD, and appropriate social support. More stakeholders in the field, including people with OCD, mental health professionals, policy makers, should be involved to enable a more inclusive and accessible environment for people with OCD.


\subsection{Technology Adoption Concerns}
The low adoption rate of technology by people with OCD revealed in our study raises concerns about the practicality of our proposed technology design opportunities. Prior research in accessibility has pointed out key factors that affect the adoption of assistive technologies, including ease of use \cite{scherer2017technology, phillips1993predictors}, trust \cite{scherer2017technology}, self-consciousness \cite{parette2004assistive, scherer2017technology}, and social acceptance \cite{shinohara2011shadow, scherer2017technology}. These factors all can potentially affect the adoption rate of suggested technology for OCD just-in-time intervention. In addition, mixed-reality is still in its early stages \cite{cummings2023psychological}, and options for consumer level physiological sensing technology is limited. Given this situation, mainstream technology such as smartphone- and smartwatch-based interventions might remain the most practical solution due to their broad user base, high social acceptance \cite{shinohara2011shadow} and increasingly powerful sensors and processors that can support our proposed intervention designs \cite{dai2021comparing, incel2023device}.

\section{Limitations}
Our research has several limitations. While our OCD participants represented diverse OCD themes, it is unclear whether our findings apply to less common OCD themes, e.g., relationship OCD. Future research should include a broader range of OCD themes to explore differences in management strategies and needs. In addition, we recruited participants with mild to moderate OCD at the time of the study. Although nine had previously experienced severe OCD symptoms, their self-management strategies and needs may vary with symptom severity. Hence, our findings should be generalized to people with severe OCD with caution. 

Although interviews with therapists enriched our understanding of treatment frameworks and common challenges in symptom management, their perspectives reflect clinical knowledge rather than lived experience. While prioritizing the views of people with OCD, the involvement of OCD professionals in our study may have influenced how we organized and interpreted participant responses. Future research should involve more people with OCD from diverse background as co-researchers to ensure that lived experience remains central throughout the research process and to counterbalance clinical assumptions. 

We acknowledge the potential compensation issue in our study. We provided reasonable compensation to both participant groups but offered OCD professionals a higher hourly rate to facilitate recruitment given their limited availability. This imbalance in compensation may raise ethical concerns. Future research should strive to provide comparable compensation for similar tasks across participant groups, based on time commitment and participation burden, to ensure fairness in the recruitment process. 

Lastly, our work is a need-finding study \cite{patnaik1999needfinding} aimed at surfacing challenges in OCD self-management and identifying technology opportunities. In the future work, we will implement and evaluate the interventions proposed in this study to validate their feasibility and effectiveness.

\section{Conclusion}
Our research contributed the first in-depth investigation to understand OCD people's specific experiences and their technology needs for just-in-time symptom management. Through an interview study with 10 participants with diverse OCD symptoms and seven OCD professionals, we identified individualized triggers and compulsions, explored properties that influence trigger intensity, and discussed how these symptoms can be recognized and tracked. We also examined the self-management strategies adopted by OCD participants, highlighting both alignment with and deviations from professional-recommended practices. Finally, we identified gaps in OCD self-management despite currently available resources, and discuss design opportunities for personalized symptom tracking, just-in-time OCD intervention, and addressing privacy and social needs that are unique to OCD.

\begin{acks}
We sincerely thank all participants who generously shared their time, experiences, and insights with us. This work would not have been possible without their willingness to contribute. We also thank Ting Zhou and Linda Zeng for their valuable input on the study design during the early stages of the project.
\end{acks}

\bibliographystyle{ACM-Reference-Format}
\bibliography{reference}

\appendix
\renewcommand\thefigure{\thesection.\arabic{figure}}    
\setcounter{figure}{0}  
\renewcommand\thetable{\thesection.\arabic{table}}    
\setcounter{table}{0}  

\newpage

\section{Appendix}
\begin{table*}
\small
\centering
\begin{tabular}{m{0.5cm}|m{5cm}|m{5cm}|m{5cm}}
\toprule
\textbf{ID} & \textbf{Obsessions} & \textbf{Representative Triggers} & \textbf{Corresponding Compulsions} \\
 \hline
\multirow{4}{*}{P1} &  & Coming home from outside & Washing hands and clothes \\
\cline{3-4}
 & Concern with contamination & Seeing dirty dishes with food residual & Cleaning the kitchen \\
 \cline{3-4}
 &  & Using trash cans & Avoiding trash cans \\
 \cline{2-4}
 & Concern with not being told the truth & No specific triggers & Seeking reassurance whether someone is telling the truth \\
 \hline
\multirow{4}{*}{P2} &  & Holding a knife in hand & Pointing knife towards herself to protect others being harmed \\
\cline{3-4}
 & Intrusive thoughts about harm occuring to herself and others & \multirow{2}{*}{Being around family or friends} & Rumination about why she had the intrusive thoughts \\
 \cline{4-4}
 &  &  & Avoiding eye contact with people who triggered her OCD \\
 \cline{2-4}
 & Intrusive thoughts about engaging in sexual activity with men in an uncomfortable way & Being around other men & Looking away from people who triggered her OCD \\
 \hline
\multirow{3}{*}{P3} &  & Touching batteries (considered dirty) & Washing hands \\
\cline{3-4}
 & Concern with contamination & Cooking raw meat & Repeatedly checking meat doneness \\
 \cline{3-4}
 &  & Being around sick people & Body scanning for symptoms \\
 \hline
\multirow{3}{*}{P4} & \multirow{2}{*}{\parbox{4.5cm}{Things she does should end in even numbers}} & \multirow{2}{*}{No specific triggers} & Counting steps when walking, \\
\cline{4-4}
 &  &  & Counting syllables when talking \\
 \cline{2-4}
 & Hair should look even on both sides & Looking at herself in the mirror & Pulling hair/eye lash so that both sides look even \\
 \hline
\multirow{2}{*}{P5} & \multirow{2}{*}{Fear of making mistakes} & \multirow{2}{*}{\parbox{4.5cm}{When worrying about something or feeling lost}} & Rumination about how she did something wrong that directly or indirectly caused the outcome \\
\cline{4-4}
 &  &  & Tightening hands to the extent that it hinders other activity \\
 \hline
\multirow{4}{*}{P6} & \multirow{3}{*}{Concern with contamination} & Accidentally touching the bathroom sink when washing hands & Washing hands \\
\cline{3-4}
 &  & Knowing some man used her bathroom & Cleaning everything in the bathroom she suspects the man touched \\
\cline{3-4}
 &  & Hearing the housekeeper washing utensils & Checking if housekeeper's hands are clean \\
 \cline{2-4}
 & Concern with missing important information in conversations & No specific triggers & Repeatedly ask the interlocutor what they just said \\
 \hline
\multirow{4}{*}{P7} & \multirow{2}{*}{Concern with contamination} & Seeing dust or stain in the kitchen & Cleaning the entire kitchen \\
\cline{3-4}
 &  & Riding public transportation & Changing clothes when back home \\
 \cline{2-4}
 & \multirow{2}{*}{\parbox{4.5cm}{Fear of hitting a pedestrian while driving without knowing}} & Driving through a pothole & \multirow{2}{*}{Checking for marks on the car} \\
  \cline{3-3}
  &  & Driving under low visibility on road & \\
 \hline
\multirow{4}{*}{P8} & \multirow{2}{*}{Concern with contamination} & \multirow{2}{*}{\parbox{4.5cm}{Touching things that multiple people have touched (e.g., door knob)}} & Washing hands \\
\cline{4-4}
 &  &  & Not using hands as much as possible \\
 \cline{2-4}
 & \multirow{2}{*}{\parbox{4.5cm}{Fear of being responsible for something terrible happening}} & Leaving home & Putting hands under the faucet to make sure no water dripping \\
 \cline{3-4}
 &  & Hearing conversations about accidents & Going back and checking if the door is locked \\
 \hline
\multirow{3}{*}{P9} & Fear of being responsible for something terrible happening & When finishing cooking & Checking if the burners are off \\
\cline{2-4}
 & Fear of being late & Before bed time & Repeatedly checking if the alarms are on \\
\cline{2-4}
  & Fear of hitting a pedestrian while driving without knowing & Driving down a residential street & Checking rear mirror \\
 \hline
\multirow{3}{*}{P10} & \multirow{2}{*}{\parbox{4.5cm}{Things should be put in specific and positions}} & No specific triggers & Making sure nothing on the table is moved \\
\cline{3-4}
 &  & Seeing baking powder on countertop & Cleaning the countertop when someone is still cooking \\
 \cline{2-4}
 & Fear of deviating from routine would cause something terrible to happen & Anything delaying his original schedule (e.g., microwave occupied by others) & Counting delayed time to make up for it at the end of the day \\
 \bottomrule
\end{tabular}
\caption{Detailed personal experience of OCD (obsessions, triggers and compulsions) reported by participants. For participants who reported more than three triggers, only representative examples are listed.
}
\label{tab:ocd_symptoms}

\end{table*}

\end{document}